\documentclass[aps,prb,reprint,showpacs,superscriptaddress]{revtex4-1}
\usepackage[dvipdfmx]{graphicx}

\usepackage{amsmath,amssymb}
\usepackage{color}
\usepackage{ulem}

\newcommand{\red}[1]{\textcolor{red}{#1}}

\usepackage{bm}

\def\vb#1{{\bm#1}}
\def\v#1{\bm{#1}}			
\def\vr{\v{r}} 					
\def\vp{\v{p}} 					

\def\p{\v{p}} 					

\def\vomega{\vb{\omega}}

\def\del{\partial}

\def\la{\langle}
\def\ra{\rangle}

\def\vs{\v{s}}

\begin{document}


\title{Valley transport driven by dynamic lattice distortion}




\author{Yuya Ominato}
\affiliation{%
Kavli Institute for Theoretical Sciences, University of Chinese Academy of Sciences, Beijing 100190, China.
}%
\author{Daigo Oue}
\affiliation{%
Kavli Institute for Theoretical Sciences, University of Chinese Academy of Sciences, Beijing 100190, China.
}%
\affiliation{%
The Blackett Laboratory, Department of Physics, Imperial College London, Prince Consort Road, Kensington, London SW7 2AZ, United Kingdom
}%
\author{Mamoru Matsuo }
\affiliation{%
Kavli Institute for Theoretical Sciences, University of Chinese Academy of Sciences, Beijing 100190, China.
}%

\affiliation{%
CAS Center for Excellence in Topological Quantum Computation, University of Chinese Academy of Sciences, Beijing 100190, China
}%
\affiliation{%
RIKEN Center for Emergent Matter Science (CEMS), Wako, Saitama 351-0198, Japan
}%
\affiliation{%
Advanced Science Research Center, Japan Atomic Energy Agency, Tokai 319-1195, Japan
}%

\date{\today}

\begin{abstract}
Angular momentum conversion between mechanical rotation and the valley degree of freedom in 2D Dirac materials is investigated theoretically. Coupling between the valley and vorticity of dynamic lattice distortions is derived by applying the $k\cdot p$ method to 2D Dirac materials with an inertial effect. Lattice strain effects are also incorporated. Valley transfer and valley-dependent carrier localization are predicted using the dynamic lattice distortions. The transport properties are found to be controllable, allowing the system to be insulating and to generate pulsed charge current. Our formalism offers a route toward mechanical manipulation of valley dynamics in 2D Dirac materials.
\end{abstract}

\maketitle 

\section{Introduction}

\begin{figure*}
\begin{center}
\includegraphics[width=1\hsize]{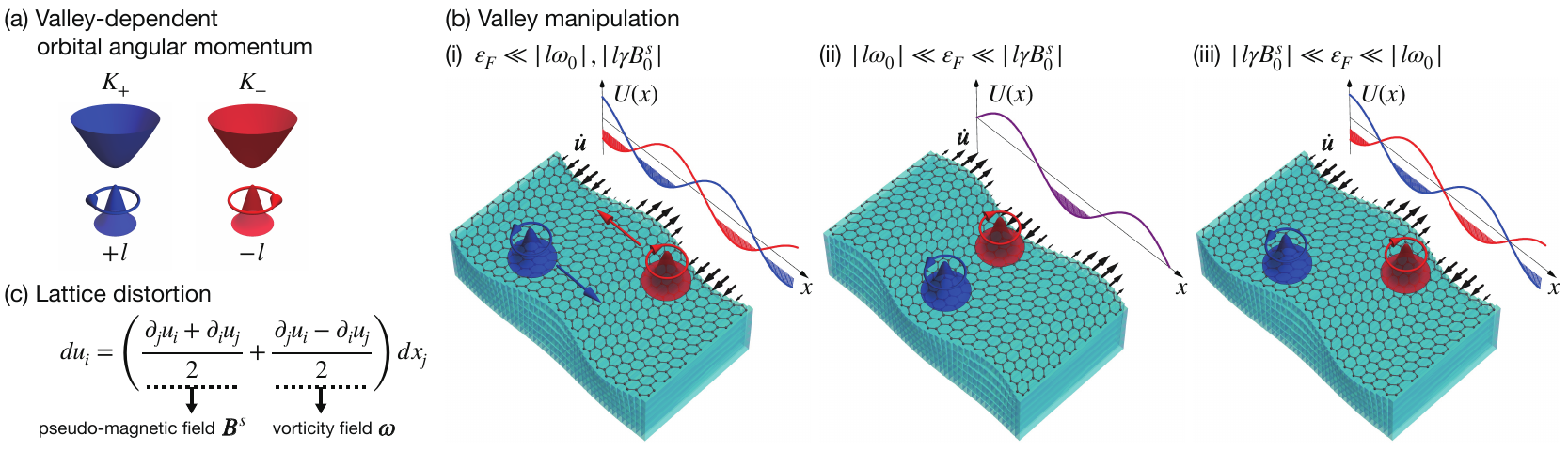}
\end{center}
\caption{
(a) Intrinsic valley-dependent orbital angular momentum for two inequivalent valleys.
(b) 2D Dirac materials on a substrate in the presence of Love-type SAWs. The sum of the pseudo-valley Zeeman coupling (pVZC), and the valley-vorticity coupling (VVC) is shown for three cases.
(i) The carriers are trapped at the bottom of $U$ and the SAW conveys the valley degree of freedom. (ii) The carriers are localized when the pVZC exceeds the Fermi energy. Both the $K_+$ and $K_-$ valleys are localized at the same location. (iii) The carriers are localized when the VVC exceeds the Fermi energy. Here, the $K_+$ and $K_-$ valley carriers are localized at different locations.
(c) The dynamic lattice distortion leads to two effective magnetic fields: the strain-induced pseudo-magnetic field $\bm{B}^s$ and the vorticity field $\bm{\omega}$.
}
\label{fig_system}
\end{figure*}

Valleytronics is an emerging field that uses the valley degrees of freedom, i.e., the local extrema of the electronic band structure, with the aim of developing innovative approaches to information processing, optoelectronic devices, and quantum computation  \cite{schaibley2016valleytronics,krasnok2018nanophotonics,vitale2018valleytronics}.
In 2D Dirac materials such as gapped graphene \cite{semenoff1984condensed,xiao2007valley} and monolayer transition-metal dichalcogenides (TMDCs) \cite{xiao2012coupled}, the electrons carry valley-dependent orbital angular momentum that originates from the Berry curvature caused by inversion symmetry breaking \cite{xiao2007valley,xiao2010berry}.
As a result, the valleys in these systems can be identified based on the intrinsic orbital angular momentum of the electrons, as illustrated in Fig.\ \ref{fig_system} (a).
This means that the valleys can be controlled using an external field coupled to the orbital angular momentum. In fact, it has been demonstrated experimentally that circularly polarized light beams \cite{cao2012valley,mak2012control,zeng2012valley,sallen2012robust,guddala2021all} and magnetic fields \cite{li2014valley,aivazian2015magnetic,macneill2015breaking,srivastava2015valley} can be used to manipulate the valley degree of freedom. Since these demonstrations, more versatile valley control methods have been desired. One candidate method taken from familiar phenomena is the gyroscopic effect, by which the angular momentum is coupled with mechanical rotation.

This gyroscopic coupling technique has been observed and used in a variety of systems. For example, gyroscopes are used to measure the gravitomagnetic field precisely in a curved space near the Earth \cite{everitt2011gravity}.  
The quantum version of the gyroscopic effect is called the gyromagnetic effect, where both the orbital and spin angular momenta are coupled to the mechanical rotation. The gyromagnetic effect was discovered during an attempt to explore the origins of magnetism \cite{EdH1915,barnett1915magnetization,barnett1935gyromagnetic} and led to the discovery of electron spin angular momentum before quantum mechanics was established.

While the original targets in these gyromagnetic experiments were limited to ferromagnetic materials, gyromagnetic coupling itself is a universal phenomenon that occurs even in nonmagnetic materials. Indeed, gyromagnetic coupling has been observed in various branches of physics, including spintronics \cite{chudo2014observation,imai2018observation,imai2019angular,takahashi2016spin,Takahashi2020giant,kazerooni2020electron,kazerooni2021electrical,kobayashi2017spin,Harii2019,okano2019nonreciprocal,kurimune2020observation,kurimune2020highly,Mori2020,tateno2020electrical,tateno2021einstein},
ultrafast demagnetization processes \cite{Dornes2019}, and quark-gluon many-body systems \cite{adamczyk2017global}.
These studies have shown that the coupling between angular momentum and rotation emerges universally in a variety of systems and continues to have a tremendous impact on a wide range of topics in fundamental physics.

In this work, we propose an alternative valley manipulation technique that uses gyroscopic coupling between the valley and the local rotational motion (or vorticity) excited by dynamic distortion in 2D Dirac materials. 
Figure\ \ref{fig_system} (b) shows schematic images of systems in which 2D Dirac materials are placed on a substrate and dynamic lattice distortion is then applied. We consider a Love wave, which is a type of horizontally polarized surface acoustic wave (SAW). This SAW leads to two effective magnetic fields: the strain-induced pseudo-magnetic field $B_z^s$ and the vorticity field $\omega_z$ as shown in Fig.\ \ref{fig_system} (c).

Both of these effective fields couple to the valley orbital angular momentum, thus allowing the valley to be manipulated using the SAW. We also consider electron doping of these 2D materials. In the presence of a standing SAW, valley transfer and dynamic valley polarization can be realized, depending on the Fermi energy $\varepsilon_F$ and the strengths of $B^s_z$ and $\omega_z$.

We also discuss the charge transport when an external DC electric field is applied to 2D Dirac materials in the presence of the SAW. We find that the conductivity is suppressed by the SAW and that a pulsed charge current is generated; this current can be used as a broadband microwave source. Our findings presented here pave the way towards valley device applications for 2D Dirac materials using the SAW devices \cite{Delsing2019-ar}.

The paper is organized as follows. The emergent gauge fields induced by the dynamic lattice distortion are introduced in Sec.\ \ref{sec_gauge_field}. The mechanism and procedure to incorporate the emergent gauge fields are explained. Finally, two kinds of the Zeeman like coupling are derived. The valley manipulation using the SAW are proposed in Sec.\ \ref{sec_valley_manipulation}. Furthermore, pulsed current generation and microwave radiation based on the proposed valley manipulation mechanism are discussed as application examples in Sec.\ \ref{sec_current} and \ref{sec_radiation}, respectively. A conclusion is given in Sec.\ \ref{sec_conclusion}.

\section{Gauge fields induced by the dynamic lattice distortion}
\label{sec_gauge_field}

We consider 2D Dirac materials in which the dynamic lattice distortion is induced by the SAW in the substrate. The dynamic lattice distortion leads to two kinds of emergent gauge fields, the velocity fields due to the inertial effect \cite{matsuo2017theory} and the strain-induced gauge field \cite{suzuura2002phonons,manes2007symmetry,vozmediano2010gauge}. They originate from the different mechanism. In the following subsections, we describe how each gauge field arises and how it is incorporated into the low-energy effective Hamiltonian of 2D Dirac materials.

\subsection{Velocity field due to the inertial effect}

In this subsection, we focus on the inertial effect. The period of the SAW is about a nanosecond and the typical lifetime of electrons would be about a picosecond, so that an adiabatic approximation is valid, where electrons adiabatically follow the dynamic lattice distortion.
As a result, the effect of the dynamic lattice distortion is incorporated as the inertial effect \cite{matsuo2017theory}. Starting from the generally covariant Dirac Lagrangian, which governs the dynamics of the spin-1/2 particle in curved spacetime, the inertial effect is incorporated into the non-relativistic Hamiltonian as a $U(1)$ potential
\begin{align}
    H=\frac{(\bm{p}+e\bm{A}^v)^2}{2m}+V(\bm{r}),
    \label{eq_non_relativistic_H}
\end{align}
where $V(\bm{r})$ is a periodic potential of a crystal and $\bm{A}^v$ is the emergent gauge field
\begin{align}
    \bm{A}^v=-\frac{m}{e}(\dot{u}_x,\dot{u}_y,\dot{u}_z),
\end{align}
with the velocity field of the lattice $\dot{\bm{u}}$. The detailed derivation is explained in the Appendix. Therefore, the inertial effect is incorporated into the low-energy effective model as conventional electromagnetic field.


Based on the discussion in the above paragraph, the low-energy effective Hamiltonian for 2D Dirac materials, e.g., gapped graphene \cite{semenoff1984condensed,xiao2007valley} and TMDCs \cite{xiao2012coupled}, is given by
\begin{align}
    H_{2D}=v(\xi\sigma_x\pi_x+\sigma_y\pi_y)+\Delta\sigma_z,
    \label{eq_kp_hamiltonian}
\end{align}
where $v$ is the velocity, $\bm{\sigma}$ are Pauli matrices that describe the pseudo spin, $\xi=\pm1$ specifies the states at the $K_+$ and $K_-$ valleys, $\bm{\pi}$ is the kinetic momentum, and $\Delta$ is an asymmetric potential breaking the inversion symmetry.
The effect of the dynamic lattice distortion can be incorporated as emergent gauge fields. We then substitute $\bm{\pi}=\bm{p}+e\bm{A}^v$ into the above.

\subsection{Strain-induced gauge field}

In addition to the velocity field, the dynamic lattice distortion also leads to the strain-induced gauge field given by \cite{suzuura2002phonons,manes2007symmetry,vozmediano2010gauge}
\begin{align}
    \xi\bm{A}^s=\xi\frac{E_0}{ev}(u_{xx}-u_{yy},-2u_{xy}),
\end{align}
with the strain tensors $u_{ij}=(\partial_j u_i+\partial_i u_j)/2$ and the material parameter $E_0$.
There are two ways to derive the strain-induced gauge field, starting from the tight-binding model and argument based on the symmetry.
In addition to the velocity field $\bm{A}^v$, the strain-induced gauge field $\xi\bm{A}^s$ has to be incorporated into the kinetic momentum $\bm{\pi}$ as $\bm{\pi}=\bm{p}+e\xi\bm{A}^s+e\bm{A}^v$.
Note here that both the velocity field and the strain-induced gauge field are induced by the dynamic lattice distortion, but it is not always the case that both fields will be generated. For example, when a rigid body rotation is considered, only the velocity field due to the inertial effect arises.

\subsection{Hamiltonian near the conduction band bottom}

Although the electronic states in 2D Dirac materials is obtained by solving Eq.\ ({\ref{eq_kp_hamiltonian}}),
it is convenient to reduce the two-band model to a one-band model to examine the physics near the conduction band bottom. Using the Schrieffer-Wolff transformation \cite{schrieffer1966relation}, the two-band model given in Eq.\ (\ref{eq_kp_hamiltonian}) can be reduced to a one-band model. The Schrieffer-Wolff transformation is a widely used technique that perturbatively incorporates the effect of off-diagonal matrix elements and systematically reduces the size of the matrix to be solved. Consequently, the effective Hamiltonian near the conduction band bottom is given by
\begin{align}
    H^{\mathrm{eff}}_{2D}
    =
    \frac{\bm{\pi}^2}{2m^\ast}
    +
    l \frac{\gamma B^s_z}{2}
    -
    \xi l\frac{\omega_z}{2},
\end{align}
where $m^\ast=\Delta/v^2$ is the effective mass, $\xi l=\xi\hbar m/m^\ast$ is the intrinsic valley-dependent orbital angular momentum \cite{xiao2007valley,xiao2010berry,koshino2010anomalous,koshino2011chiral},
$\gamma=e/m$ is the gyromagnetic ratio, $B^s_z=\partial_xA^s_y-\partial_yA^s_x$ is the strain-induced pseudo-magnetic field, and $\omega_z=\partial_x\dot{u}_y-\partial_y\dot{u}_x$ is the vorticity field.
The first term is the conventional kinetic energy term with the effective mass $m^\ast$.
The second term describes the coupling between the valley magnetic moment and the pseudo-magnetic field, which we call the pseudo valley Zeeman coupling (pVZC).
The second term is independent of the valley index because the strain field preserves the time-reversal symmetry.
The third term describes the coupling between the valley orbital angular momentum and the vorticity field, which we call the valley-vorticity coupling (VVC).
The VVC is similar to the valley Zeeman coupling, which describes the coupling between the valley magnetic moment and a magnetic field \cite{li2014valley,aivazian2015magnetic,macneill2015breaking,srivastava2015valley}.
The derivation of the VVC is one of the main results reported in this paper.
Note here that the acoustoelectric effect in the TMDC was discussed previously in a system similar to our setup \cite{kalameitsev2019valley,Sonowal2020-ul}; however, the coupling between the dynamic lattice distortion and the valley orbital angular momentum is discussed for the first time in this work.

\section{Valley manipulation using SAW}
\label{sec_valley_manipulation}

In our setup, the Love wave is excited in the substrate by an external force, which leads to the dynamic lattice distortion in 2D Dirac materials. The detailed properties of the Love wave are explained in the Appendix. The lattice displacement vector in 2D Dirac materials is given by
\begin{align}
    \bm{u}=(0,u_0\sin(kx)\sin(\omega t),0),
\end{align}
which means that the vorticity field and the pseudo-magnetic field are given by
\begin{align}
\bm{\omega}=(0,0,\omega_0\cos(kx)\cos(\omega t))
\end{align}
with $\omega_0=u_0\omega^2/c_s$ and
\begin{align}
    \bm{B}^s=(0,0,B_0^s\sin(kx)\sin(\omega t))
\end{align}
with $B_0^s=\omega_0E_0/evc_s$, respectively.
As shown in the Appendix, the dispersion relation is approximately linear, so that we use the dispersion relation $\omega=c_s k$, where $c_s$ is the SAW velocity.
The sum of the pVZC and the VVC can then be treated as a periodic potential
\begin{align}
    U=l\frac{\omega_0}{2}
    \left[
        R\sin(kx)\sin(\omega t)
        -
        \xi\cos(kx)\cos(\omega t)
    \right],
    \label{eq_periodic_potential}
\end{align}
where the dimensionless material parameter $R$ is introduced
\begin{align}
    R=\frac{\gamma E_0}{evc_s}.
\end{align}
Different valley manipulations are possible, depending on the value of $R$.
Note that the dynamic lattice distortion induced here is not directly related to the phonon modes in the 2D Dirac materials. We expect that the effect due to the pVZC and the VVC can be detected even in the presence of the thermally activated phonon modes. Indeed, phenomena related to our proposal, such as strain induced Landau level formation in graphene \cite{nigge2019room} and spin manipulation using the SAW device \cite{weiler2011elastically,kobayashi2017spin,kurimune2020observation}, have been observed at room temperature.

We consider three cases showing characteristic carrier distribution in the electron-doped system.
\begin{itemize}
\item (i) When the Fermi energy is sufficiently smaller than both the pVZC and the VVC (i.e., when $\varepsilon_F\ll|l\gamma B_0^s|,|l\omega_0|$), the carriers are trapped at the bottom of potential and are transferred in the opposite direction for each valley, as shown in Fig.\ \ref{fig_system} (b)(i).
This is a promising candidate mechanism to convey the valley information.
This case can be realized using any value of $R$.
\item (ii) When the Fermi energy is sufficiently larger than the VVC but smaller than the pVZC (i.e., when $|l\omega_0|\ll\varepsilon_F\ll|l\gamma B_0^s|$), the carriers are then localized at the time $\omega t=(N+1/2)\pi,\ (N=0,\pm1,\cdots)$, as shown in Fig\ \ref{fig_system} (b)(ii). In this case, the spatial distribution of the carriers is almost independent of the valley.
This case can be realized when $R\gg1$.
\item (iii) When the Fermi energy is sufficiently larger than the pVZC but smaller than the VVC
(i.e., when $|l\gamma B_0^s|\ll\varepsilon_F\ll|l\omega_0|$),
the carriers are then localized at the time $\omega t = N\pi,\ (N=0,\pm1,\cdots)$, as shown in Fig.\ \ref{fig_system} (b)(iii).
In contrast to the second case, each valley's carrier is localized in a different location separated by $\pi/q$.
This case can be realized when $R\ll1$.
\end{itemize}

The valley transfer shown in Fig.\ \ref{fig_system} (b)(i) and the carrier localizations shown in Fig.\ \ref{fig_system} (b)(ii) and (iii) are detectable experimentally using time-resolved optical diffraction measurements.
Because the carrier density pattern is periodic, diffraction patterns will appear under coherent illumination. The periods shown in Fig.\ \ref{fig_system} (b)(ii) and (iii) are different and thus the corresponding diffraction patterns are also different.
Furthermore, the valley dependence of the carriers is also detectable using Kerr rotation microscopy, because the valleys carry an orbital magnetic moment \cite{lee2016electrical}.
The experimental signatures are listed in Table. \ref{table_exp}.

\begin{table}[t]
\caption{\label{table_exp}
Experimental signatures to detect the valley manipulations discussed in this work.
}
\begin{ruledtabular}
\begin{tabular}{cccc}
    Case &
    (i)   &
    (ii)  &
    (iii) \\
    \hline\hline
    Optical diffraction&
    $\checkmark$ &
    $\checkmark$ &
    $\checkmark$ \\
    \hline
    Kerr rotation&
    $\checkmark$ &
    -- &
    $\checkmark$ \\
    \hline
    Pulsed current&
    -- &
    $\checkmark$ &
    $\checkmark$ \\
\end{tabular}
\end{ruledtabular}
\end{table}

Next, we estimate the strengths of the vorticity field and the pseudo-magnetic field.
Using the parameters of graphene from the previous study \cite{suzuura2002phonons}, the relationship for the strengths of the vorticity field and the pseudo-magnetic field is given as $\omega_0/\gamma\sim10^{-3}\times B^s_0$, and when the parameters of $\mathrm{MoS}_2$ from the preious study \cite{rostami2015theory} are used, the relationship is given as $\omega_0/\gamma\sim B_0^s$.
The experimentally feasible parameters for the SAW are $u_0\approx  100\ \mathrm{pm}$, $c_t\approx 10^3\ \mathrm{m/s}$, and $\omega\approx 30\ \mathrm{GHz}$, meaning the strength of the vorticity field can be estimated to be $\omega_0/\gamma\sim0.01\ \mathrm{T}$.


\section{Manipulation of transport properties}
\label{sec_current}

In addition to the spatial distribution of the carriers, the carrier transport properties are also modulated by the SAW.
Here, we discuss their longitudinal transport properties along the $x$ direction in the three cases described above based on the semiclassical Boltzmann transport theory.
In case (i),
the carriers are trapped at the bottom of the potential $U$ at any time, which means that the system becomes insulating.
In cases (ii) and (iii), conversely, the system undergoes repeated carrier localization and delocalization when the amplitude of $U$ exceeds and falls below $\varepsilon_F$, respectively, because of the time dependence of $U$.
To investigate the longitudinal transport properties in the latter cases, we use semiclassical Boltzmann transport theory.
In the following discussion, we focus on the second case and omit the VVC for simplicity, which corresponds to the assumption that $R\gg1$.
This simplification does not affect the main results given below because the effect of the VVC is negligible due to the condition $|l\omega_0|\ll\varepsilon_F$.
As shown in Eq.\ (\ref{eq_periodic_potential}), although there is a phase difference between the pVZC and the VVC, their qualitative effects on the longitudinal transport properties are almost the same.
Therefore, the results are also applicable to the third case.

The semiclassical Boltzmann kinetic equation is given by
\begin{align}
    \frac{\partial f_{\xi}}{\partial t}
    +
    \dot{\vr}\cdot\frac{\partial f_{\xi}}{\partial \vr}
    +
    \dot{\vp}\cdot\frac{\partial f_{\xi}}{\partial\vp}
    =
    \left(
    \frac{\partial f_{\xi}}{\partial t}
    \right)_{\mathrm{coll}},
\end{align}
where $f_{\xi}$ is the distribution function for the valleys $K_+$ and $K_-$, and the equations of motion are given by
\begin{align}
    \dot{\vr}
    &=
    \frac{\partial\varepsilon_{\xi,\vp}}{\partial\vp}
    -\dot{\vp}\times\bm{\Omega}_{\xi}, \label{eq_rdot} \\
    \dot{\vp}
    &=
    -e\bm{E}
    +m\dot{\vr}\times\bm{\omega}
    -e\dot{\vr}\times\xi\bm{B}^s, \label{eq_kdot}
\end{align}
where $\varepsilon_{\xi,\vp}$ is the energy band in the presence of the pVZC, $\bm{\Omega}_{\xi}=\hbar\nabla_{\vp}\times i\la u_{\xi,\vp}|\nabla_{\vp}|u_{\xi,\vp}\ra$ is the Berry curvature for each valley, and $|u_{\xi,\vp}\ra$ is the periodic part of the Bloch wave function.
The second term in Eq.\ (\ref{eq_rdot}) is the anomalous velocity \cite{xiao2010berry}, the second term in Eq.\ (\ref{eq_kdot}) is the Coriolis force that originates from the vorticity field of the lattice \cite{matsuo2017theory}, and the third term in Eq.\ (\ref{eq_kdot}) is the valley-dependent Lorentz force that originates from the pseudo-magnetic field.

Next, we consider the charge transport along the $x$-direction in the presence of the external electric field $\bm{E}=(E_x,0,0)$.
We use the following relaxation time approximation:
\begin{align}
    \left(
    \frac{\partial f_{\xi}}{\partial t}
    \right)_{\mathrm{coll}}
    =
    -\frac{f_{\xi}-f^{(0)}_{\xi}}{\tau},
\end{align}
where $\tau$ is the valley independent relaxation time and $f^{(0)}_{\xi}=1/(1+\exp[(\varepsilon_{\xi,\vp}-\varepsilon_F)/k_BT])$ is the Fermi distribution function.
The above expression of the relaxation time approximation is based on the three assumptions. First, the valley quantum number is treated as a conserved quantity. This assumption is reasonable if the sample is sufficiently clean that the effect of the atomic scale scatterers is negligible and there is no intervalley scattering process. In addition, the SAW rarely leads to intervalley scattering because the wavelength of the SAW is much larger than the lattice constant. 
Next, the relaxation time approximation is valid even in the presence of the time-dependent potential because the SAW period is much larger than the lifetime of the electron. In our setup, the SAW period is about a nanosecond, and the lifetime of the electron would be about a picosecond.
Finally, the relaxation time is independent of the valley quantum number. This assumption is reasonable if the scatterers have no characteristic feature that breaks the valley degeneracy.

Assuming a weak vorticity field and a weak pseudo-magnetic field (i.e., $\omega_0 m|\Omega_{\xi,z}|\ll1$, $\omega_0\tau\ll1$, and $\omega_c\tau\ll1$ with $\omega_c=eB^s_0/m^\ast$), the anomalous velocity, the Coriolis force, and the Lorentz force then only give corrections for the longitudinal conductivity and have almost no effect on the qualitative behavior of the longitudinal conductivity.
Based on this assumption, we can omit these three terms.
Note here that there is no Hall voltage in the current setup because the net vorticity field is zero.
The SAW dynamics are much slower than the dynamics of the electrons, which means that the adiabatic approximation is valid and it can be assumed that the steady state is achieved at each moment.
Finally, the charge current can be calculated by deriving the steady state of the distribution function at each moment and using snapshot energy bands.
We define $\delta f_{\xi}$ as $f_\xi=f^{(0)}_\xi+\delta f_{\xi}$, and $\delta f_{\xi}$ is given by
\begin{align}
    \delta f_{\xi}
    =
    \tau e E_xv_x
    \left(
        \frac{\partial f_{\xi}}{\partial \varepsilon_{\xi,\vp}}
    \right),
\end{align}
where $v_x=\partial \varepsilon_{\xi,\vp}/\partial p_x$.
Using $\delta f_{\xi}$, the charge current at $T=0$ can be given by
\begin{align}
    j_{x}
    =
    \sum_{\xi=\pm1}
    e^2\tau E_x
    \int\frac{d\vp}{(2\pi\hbar)^2}v_x^2
    \delta(\varepsilon_F-\varepsilon_{\xi,\vp}).
    \label{eq_charge_current}
\end{align}

\begin{figure}[t]
\begin{center}
\includegraphics[width=1\hsize]{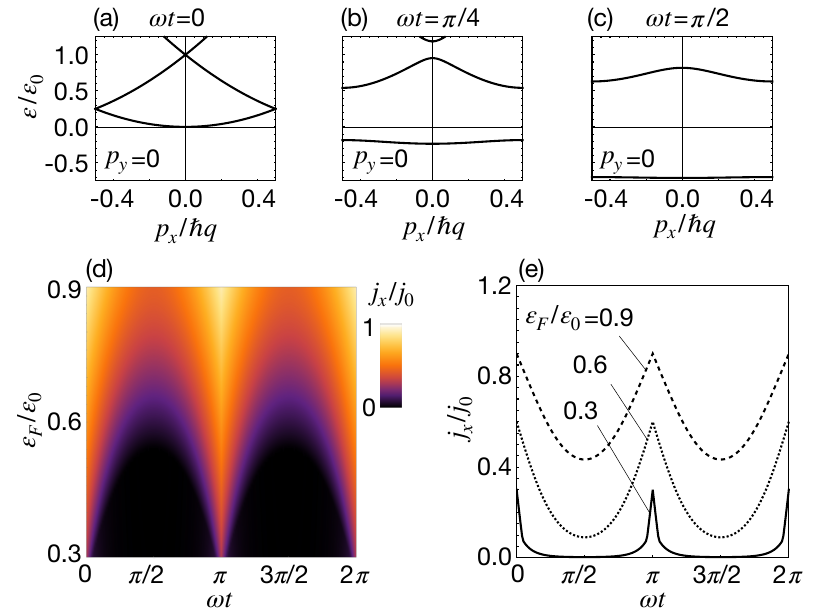}
\end{center}
\caption{
(a)-(c) Snapshot energy bands at several values of $\omega t$.
(d) Charge current as a function of both $\varepsilon_F$ and $\omega t$.
(e) Charge current as a function of $\omega t$ at several values of $\varepsilon_F$. We set $|l\gamma B^s_0|/2=1.5\varepsilon_0$, where $\varepsilon_0=\hbar^2 q^2/2m^\ast$. The unit of $j_x$ is given by $j_0=(n_0e^2\tau/m^\ast)E_x$ with carrier density $n_0=\varepsilon_0m^\ast/2\pi\hbar^2$.
}
\label{fig_conductivity}
\end{figure}

Figure \ref{fig_conductivity} (a)-(c) show snapshot energy bands as a function of $p_x$ with a fixed $p_y=0$ at several values of $\omega t$.
The pVZC is treated as a spatially periodic potential in each case, which means that the energy gap opens and the group velocity decreases as the amplitude $B_z^s$ increases.
Therefore, the charge current is suppressed by the SAW.
Figure \ref{fig_conductivity} (d) shows the charge current as a function of $\omega t$ and the Fermi energy $\varepsilon_F$.
At a fixed $\varepsilon_F$, the charge current changes with the period $\pi$ and reaches a maximum value at $\omega t=N\pi,\ (N=0,\pm1,\cdots)$.
The charge current is almost zero around $\omega t=(N+1/2)\pi,\ (N=0,\pm1,\cdots)$ when the Fermi energy is sufficiently small in comparison to the pVZC.
Figure \ref{fig_conductivity} (e) shows the charge current as a function of $\omega t$ at several fixed $\varepsilon_F$. One can see that the sharp pulsed current is generated when $\varepsilon_F\ll|l\gamma B_0^s|$.
At a finite temperature, the pulsed current is subject to thermal broadening. The sharp pulsed current can still be generated in a condition that the thermal broadening energy is much smaller than the pVZC (i.e., $k_BT\ll|l\gamma B_0^s|$). Using the parameters for graphene with energy gap $\Delta=0.1\ \mathrm{eV}$ \cite{zhou2007substrate,zhou2008origin}, the pVZC is estimated as $|l\gamma B_0^s/2|\sim\ 30\mathrm{meV}$, which is the same order of the thermal broadening energy at room temperature.
Therefore, the condition to generate the sharp pulsed current is expected to be experimentally feasible.
Within the quadratic approximation, the Fermi energy $\varepsilon_F$ is proportional to the carrier density $n$ (i.e., $\varepsilon_F=2\pi\hbar^2n/m^\ast$), so that one can replace the Fermi energy in Fig.\ \ref{fig_conductivity} with the carrier density by multiplying the factor $m^\ast/2\pi\hbar^2$.

To summarize the discussion of the transport properties above, when the Fermi energy is sufficiently smaller than both the pVZC and the VVC, the system becomes insulating, but when the Fermi energy is larger than either the pVZC or the VVC, the conductivity is then suppressed and a pulsed current can be generated, as listed in Table. \ref{table_exp}.

\section{Broadband microwave sources}
\label{sec_radiation}

\begin{figure}[t]
\begin{center}
\includegraphics[width=1\hsize]{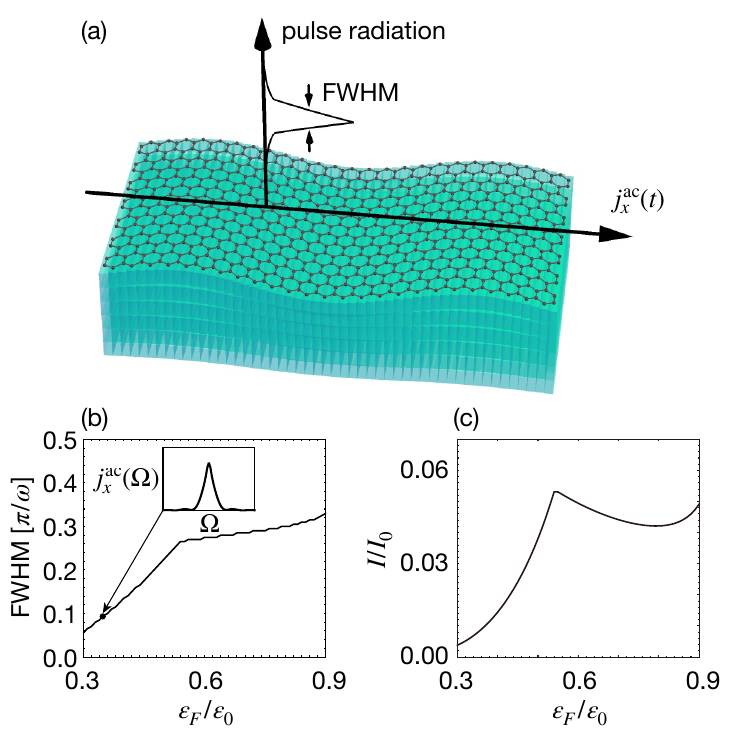}
\end{center}
\caption{
(a) Schematic image of the microwave radiation produced by $j_x^{\mathrm{ac}}(t)$. (b) Pulse FWHM as a function of the Fermi energy. The inset shows the Fourier components of the pulse at $\varepsilon_F/\varepsilon_0=0.35$.
(c) Averaged radiation intensity as a function of the Fermi energy. The unit is set as $I_0=\frac{Z_0}{2}\frac{\sqrt{\epsilon_+}}{\sqrt{\epsilon_+}+\sqrt{\epsilon_-}}j_0^2$.
}
\label{fig_radiation}
\end{figure}

The charge current obtained above can be divided into an AC component and a DC component using the form $j_x(t)=j^{\mathrm{ac}}_x(t)+j^{\mathrm{dc}}_x$, with $j^{\mathrm{dc}}_x=j_x(\pi/2\omega)$.
The DC component $j^{\mathrm{dc}}_x$ corresponds to the minimum value of the charge current and becomes finite when $\varepsilon_F/\varepsilon_0\gtrsim0.54$.
The oscillating current $j^{\mathrm{ac}}_x(t)$ can be used for broadband and tunable microwave sources, which is useful for device applications \cite{Hu2021}.
A schematic image of the microwave radiation in our setup is shown in Fig.\ \ref{fig_radiation}(a).

Figure\ \ref{fig_radiation} (b) shows the Fermi energy dependence of the full width at half maximum (FWHM) value of the pulse.
When the Fermi energy increases, the FWHM also increases, and the range of Fourier components decreases.
There is a kink structure of the FWHM at $\varepsilon_F/\varepsilon_0\approx0.54$ because $j^{\mathrm{dc}}_x$ becomes finite and increases above this point.

The averaged radiation intensity of the microwave in the upper half space $I$ is given by
\begin{align}
    I
    =
    \frac{Z_0}{2}
    \frac{\sqrt{\epsilon_+}}{\sqrt{\epsilon_+}+\sqrt{\epsilon_-}}
    \int^{\pi/\omega}_0\frac{d(\omega t)}{\pi}
    \left[j^{\mathrm{ac}}_x(t)\right]^2,
    \label{radiation_intensity}
\end{align}
where $Z_0\approx377\ \Omega$ is the characteristic impedance of free space, and $\epsilon_+$ and $\epsilon_-$ are permitivity of free space and substrate, respectively. Figure\ \ref{fig_radiation} (c) shows the averaged intensity as a function of the Fermi energy.
The averaged radiation intensity becomes a local maximum at $\varepsilon_F/\varepsilon_0\approx0.54$ above which $j^{\mathrm{dc}}_x$ becomes finite.
Setting the charge current as $j_0\sim1\ \mathrm{A}/\mathrm{m}$ and the permitivity as $\epsilon_\pm\sim1$, the averaged radiation intensity is estimated as $I\sim0.1\ \mathrm{mW}/\mathrm{cm}^2$ in the energy range considered here.

\section{Conclusion}
\label{sec_conclusion}

In this work, we have presented an investigation of the low-energy effective Hamiltonian of 2D Dirac materials in the presence of the dynamic lattice distortions.
We have derived the valley-vorticity coupling (VVC), which is regarded as an inertial effect.
In addition to the VVC, we also incorporated the strain-induced pseudo-magnetic field into the analysis.
We have proposed mechanisms for both valley transfer and valley-dependent carrier localization using an SAW.

We have also investigated the electronic transport properties in the presence of the SAW.
The conductivity along the direction of propagation of the SAW is suppressed by the dynamic lattice distortions. Periodic modulation of the conductivity results in a pulsed charge current when an electrostatic field is applied along the SAW.
This pulsed charge current act as a broadband microwave source and can be used for optoelectronic device applications.
These results pave the way toward the realization of next-generation valleytronics device applications using the SAW devices.

{\it Acknowledgements.---}
The authors are grateful to Y. Nozaki and A. Yamakage providing valuable comments.
D.O.~is funded by the President's PhD Scholarships at Imperial College London.
This work was supported by the Priority Program of Chinese Academy of Sciences under Grant No. XDB28000000, and by JSPS KAKENHI for Grants (Nos. JP20H01863 and JP21H04565) from MEXT, Japan. 

\section*{Appendix}

\renewcommand{\theequation}{A.\arabic{equation}}
\setcounter{equation}{0}

\subsection{Inertial effect due to dynamic lattice distorsions}

Let us consider the inertial effects due to dynamic lattice distortions such as elastic motion in the presence of SAWs. Previous studies have shown that the strain field caused by these lattice distortions leads to an strain-induced gauge field \cite{suzuura2002phonons,manes2007symmetry,vozmediano2010gauge}.
In the following subsequent discussion, we omit the the strain-induced gauge field and focus on the inertial effect.

The generally covariant Dirac Lagrangian, which governs the dynamics of the spin-1/2 particle in curved spacetime, is \cite{brill1957interaction,hehl1976general}
\begin{align}
    \mathcal{L} = \bar{\Psi}\left[i \gamma^{\hat{a}}e_{\hat{a}}{}^\mu(x)  \left(p_{\mu }-\mathcal{A}_{\mu}  \right)  -mc  \right] \Psi,\label{gDirac}
\end{align}
where $m$ and $c$ represent the mass of an electron and the speed of light, respectively.  
The spin connection $\mathcal{A}_{\mu}$ is given by $\mathcal{A}_\mu = i\hbar \omega_{\hat{a}\hat{b}\mu} [\gamma^{\hat{a}},\gamma^{\hat{b}}]/8$, where $\hbar$ represents the Planck's constant, $\omega_{\hat{a}\hat{b}} $ is related to the tetrad $e_{\hat{a}}$ as $ de_{\hat{a}} = \omega_{\hat{a}}{}^{\hat{b}} \wedge e_{\hat{b}}$, and $\gamma^{\hat{a}}$ is the gamma matrix. Note that $\mu$ in Eq.~\eqref{gDirac} represents the vector indices in curved spacetime and the hatted indices $\hat{a}, \hat{b}$ represent the local Lorentz indices. 

We can describe the relationship between a local rest frame on the lattice and an inertial frame in terms of the lattice displacement vector $\bm{u}$ and the lattice velocity field $\dot{\bm{u}}(x) \,\, (|\dot{\bm{u}}|/c \ll 1)$, $d\v{r}'=d\v{r} + \dot{\v{u}}(x)dt$.
We can explicitly write the tetrad
$e_{\hat{0}}{}^0 =1, e_{\hat{0}}{}^i = - \dot{u}^i/c, e_{\hat{j}}{}^0 =0, e_{\hat{j}}{}^i = \delta^i_j. $
As a result, the generally covariant Dirac Lagrangian (\ref{gDirac}) leads to the Dirac Hamiltonian in the local rest frame \cite{matsuo2017theory}
\begin{eqnarray}
H_{\rm D} &=& \beta mc^{2} + (c \vb{\alpha} -\dot{\bm{u}}(x))\cdot  \p 
	- \vb{\Sigma} \cdot \frac{\vb{\omega}(x)}{2},  \label{Hlr}
\end{eqnarray}
where 
$ \beta = \gamma^0 $ and $ \alpha_i = \gamma^0\gamma^i$ are the Dirac matrices, and $\Sigma_a = \frac{\hbar}{2} \epsilon_{abc}[\gamma^b,\gamma^c] $ is the spin operator, and $\vomega (x) = \nabla \times \dot{\bm{u}}(x)$ is the vorticity field of the lattice distortion. 
Here, two inertial effects on the spinor field due to the lattice velocity field $\dot{\bm{u}}(x)$ are included in the Hamiltonian. First, the velocity operator of the Dirac spinor in the inertial frame $c\vb{\alpha}$ is replaced with $c\vb{\alpha} - \dot{\bm{u}}(x)$. Second, we include the spin-vorticity coupling $- \vb{\Sigma} \cdot \vb{\omega}(x)/2$.

The lowest order of the Foldy-Wouthuysen-Tani transformation \cite{foldy1950dirac,tani1951connection} for Eq.\ (\ref{Hlr}) leads to the Pauli-Schr\"odinger equation for an electron in the local rest frame on the lattice
\begin{eqnarray}
i\hbar \frac{\del \psi}{\del t}=H' \psi,\,\,  
H' = \frac{(\p + e\bm{A}^v(x) )^{2}}{2m} 
	-\v{s} \cdot \frac{\vb{\omega}(x)}{2},\label{Hnr}
\end{eqnarray}
where $\v{s}$ is the spin angular momentum of the electron, which obeys the commutation relation $[s_i,s_j] =i\hbar  \epsilon_{ijk}s_k$.
This result shows that the inertial effects can be introduced into the conventional Hamiltonian in an inertial frame via the following two emergent gauge fields: the U(1) potential
\begin{align}
    \bm{A}^v(x)
    =-\frac{m}{e}\dot{\bm{u}}(x)
    =-\frac{m}{e}\left(\dot{u}_x(x),\dot{u}_y(x),\dot{u}_z(x)\right),
\end{align}
and the spin-vorticity coupling, or the SU(2) scalar potential, $- \vs \cdot \vb{\omega}(x)/2$.
Consequently, the inertial effect of the U(1) potential $\bm{A}^v$ is incorporated into the low energy effective Hamiltonian by the Peierls substitution $\bm{p}\to\bm{p}+e\bm{A}^v$ as the conventional electromagnetic field.
In the main text, we omit the spin-vorticity coupling, based on the assumption that its effect is small when compared with the orbital effect that originates from the Peierls substitution. This assumption is valid for the 2D Dirac materials because the intrinsic orbital angular momentum is larger than the spin angular momentum.
Adding the periodic potential of the lattice, which is omitted in the above discussion, gives the non-relativistic Hamiltonian Eq.\ (\ref{eq_non_relativistic_H}) used in the main text.

\renewcommand{\theequation}{B.\arabic{equation}}
\setcounter{equation}{0}

\subsection{Surface acoustic wave: Love wave}
\label{sec_SAW_Love_wave}

In this section, we explain the Love wave, a horizontally polarized (transverse) surface acoustic wave.
The equation of motion for an isotropic elastic body is given by \cite{love1911,charles1974,landau1986}
\begin{align}
    \frac{\partial^2 \bm{u}}{\partial t^2}
    =
    c_t^2\Delta\bm{u}+(c_\ell^2-c_t^2)\nabla(\nabla\cdot\bm{u}),
    \label{eq_eom_elastic_body}
\end{align}
where $\bm{u}$ is the displacement vector, and $c_t$ and $c_\ell$ are the velocities of the transverse and longitudinal waves, respectively.
In the following, we focus on transverse waves oscillate in the $y$-direction (i.e., $\bm{u}=(0,u_y,0)$). From the transversality condition (i.e., $\nabla\cdot\bm{u}=0$), the second term on the right-hand side vanishes.

\begin{figure}
    \centering
    \includegraphics[width=1\hsize]{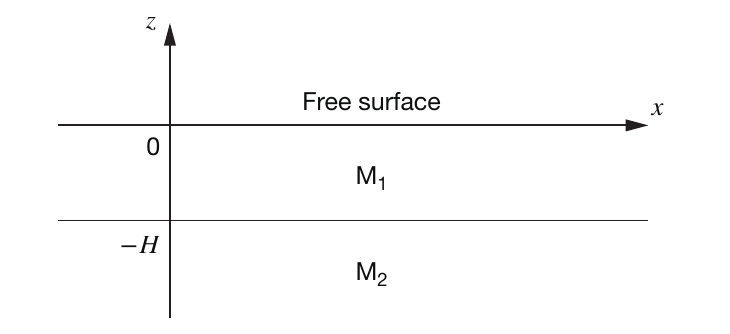}
    \caption{The slab medium $M_1$ of thickness $H$ is stacked on the semi-infinite medium $M_2$.}
    \label{fig_M1_M2}
\end{figure}

We consider a slab medium $M_1$ with a thickness of $H$ on a semi-infinite medium $M_2$, as shown in Fig.~\ref{fig_M1_M2}.
We assume solutions of Eq.~(\ref{eq_eom_elastic_body}) of the form
\begin{align}
    &u_a(x,z,t)=h_a(z)e^{i(kx-\omega t)},
    \label{eq_eom_solution}
\end{align}
where $u_a(x,z,t)~(a=1,2)$ is the $y$-component of the displacement vector in medium $M_a$.
Substituting Eq.~(\ref{eq_eom_solution}) into Eq.~(\ref{eq_eom_elastic_body}), we obtain the differential equation for $h_a(z)$
\begin{align}
    \left(
        k^2-\frac{\omega^2}{c_a^2}
    \right)
    h_a(z)
    =
    \frac{\partial^2 h_a(z)}{\partial z^2},
\end{align}
where $c_a$ is the velocity of transverse waves in each medium.
Since we consider surface waves, we assume a solution which vanishes at negative infinity (i.e., $\lim_{z\to-\infty}h_2(z)=0$).
Therefore, the solutions $h_a(z)$ are written as
\begin{align}
    &h_1(z)=Ae^{iqz}+Be^{-iqz}, \label{eq_h1} \\
    &h_2(z)=Ce^{\kappa z}, \label{eq_h2}
\end{align}
where $q$ and $\kappa$ are positive numbers given by
\begin{align}
    &q=\sqrt{\frac{\omega^2}{c_1^2}-k^2},     \label{eq_q} \\
    &\kappa=\sqrt{k^2-\frac{\omega^2}{c_2^2}} \label{eq_kappa}.
\end{align}
The velocities $c_1$ and $c_2$ have to satisfy $c_1<c_2$ since both $q$ and $\kappa$ are positive numbers.

\begin{figure}[t]
\begin{center}
\includegraphics[width=1\hsize]{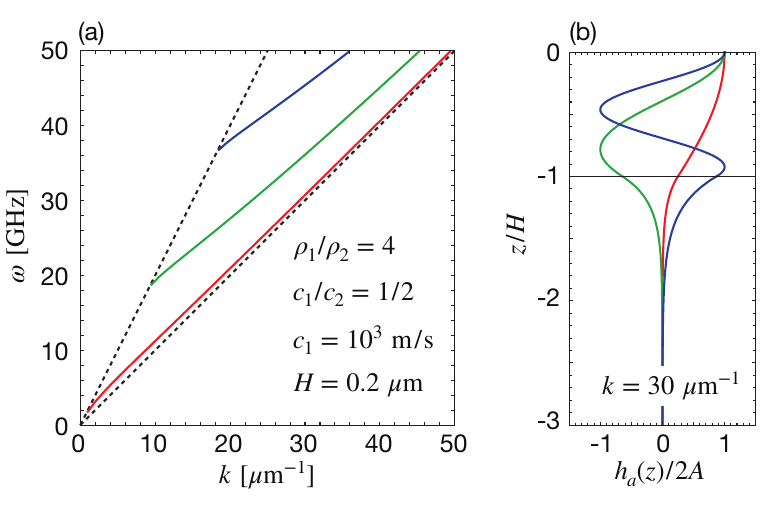}
\end{center}
\caption{
(a) The red, green, and blue curves represent the dispersion relation of the Love waves. The dotted lines are $\omega=c_1k$ and $\omega=c_2k$. We set the parameters as indicated in the figure legends.
(b) The amplitude of the Love wave $h_a(z)$ at $k=30\mu\mathrm{m}^{-1}$, where $a=1$ for $-1<z<0$ and $a=2$ for $z<-1$. The red, green, and blue curves correspond to the red, green, and blue branches in the left panel, respectively.
}
\label{fig_Love_wave}
\end{figure}

The values of $q$ and $\kappa$ are determined by finding solutions under appropriate boundary conditions.
Here, we apply the free surface boundary condition at $z=0$,
where the stress on the surface vanishes,
and the continuities of displacement and stress at the boundary between $M_1$ and $M_2$:
\begin{align}
    &\frac{\partial h_1(z)}{\partial z}\Big|_{z=0}
    =0, \\
    &h_1(-H)
    =h_2(-H), \\
    &\rho_1 c_1^2\frac{\partial h_1(z)}{\partial z}\Big|_{z=-H}
    =
    \rho_2 c_2^2\frac{\partial h_2(z)}{\partial z}\Big|_{z=-H},
\end{align}
where $\rho_a$ is the density of $M_a$.
Substituting Eq.~(\ref{eq_h1}) and (\ref{eq_h2}) into the above equations, we obtain an equation system for the coefficients $A,B,$ and $C$,
\begin{align}
    &A-B=0, \\
    &Ae^{-iqH}+Be^{iqH}=Ce^{-\kappa H}, \\
    &i\rho_1c_1^2q
    \left(
        Ae^{-iqH}-Be^{iqH}
    \right)
    =\rho_2c_2^2\kappa Ce^{-\kappa H}.
\end{align}
To obtain nontrivial solutions, we should have 
\begin{align}
    \frac{\kappa}{q}
    =
    \frac{\rho_1c_1^2}{\rho_2c_2^2}
    \tan\left(qH\right).
    \label{eq_BC}
\end{align}
Substituting Eq.~(\ref{eq_q}) and (\ref{eq_kappa}) into Eq.~(\ref{eq_BC}) to find $\omega$ and $k$ that satisfy the boundary conditions, we obtain the dispersion relation.
Figure \ref{fig_Love_wave} (a) and (b) show the dispersion relation and the amplitude of the Love waves, respectively.
The phase velocity of the Love waves $c_s$ is also obtained from the dispersion relation.
Using the obtained displacement $u_a$ and phase velocity $c_s$, the intensity of the Love waves $I_s$ is given by
\begin{align}
    I_s
    =
    \int^0_{-H}dz\frac{1}{2}c_s\rho_1|\dot{u}_1|^2
    +
    \int^{-H}_{-\infty}dz\frac{1}{2}c_s\rho_2|\dot{u}_2|^2.
\end{align}


The 2D Dirac material follows the excited Love wave on the substrate,
which dynamically distorts the lattice of the 2D material.
By superimposing of the traveling-wave-type solutions~(\ref{eq_eom_solution}),
we can get the dynamic lattice distortion of the standing-wave type,
which is employed in the main text.

\renewcommand{\theequation}{C.\arabic{equation}}
\setcounter{equation}{0}

\subsection{Microwave radiation}

In our setup, we have an oscillating electric current in the $x$-direction;
hence,
the magnetic (electric) field emitted by the current should oscillate in the $y$- ($x$-) direction.
Integrating Maxwell's equations gives the electric field continuity and the magnetic field discontinuity at boundaries.
In the frequency domain,
we can write
\begin{align}
&\lim_{\delta h \rightarrow 0}
\big[E_x(z,\omega)\big]_{z=-\delta h}^{z=+\delta h}
= 0,
\label{eq:electric-bc}
\\
&\lim_{\delta h \rightarrow 0}
\big[H_y(z,\omega)\big]_{z=-\delta h}^{z=+\delta h}
= -J_x(\omega),
\label{eq:magnetic-bc}
\end{align}
where the electric current on the 2D material $J$ is given in $\mathrm{[A/m]}$.
We have the current but no incoming electromagnetic waves.
Only outgoing waves are present, e.g.,
\begin{align}
    E_x(z,\omega)
    &= 
    \begin{cases}
        E_x^+(\omega) e^{+i (\omega/v_+) z} & z > 0,
        \\
        E_x^-(\omega) e^{-i (\omega/v_-) z} & z < 0,
    \end{cases}
\end{align}
where $v_\pm=c/\sqrt{\epsilon_\pm}$ is the speed of light in each medium with each permittivity $\epsilon_\pm$.
The magnetic field is associated with the electric field via the characteristic impedance \cite{kong1986electromagnetic},
\begin{align}
    E_x^\pm(\omega)
    = \pm \frac{Z_0}{\sqrt{\epsilon_\pm}}H_y^\pm (\omega),
\end{align}
where $Z_0 \equiv \sqrt{\mu_0/\epsilon_0} \approx 377\ \Omega$ is the characteristic impedance of free space.
We rearrange Eqs.~(\ref{eq:electric-bc}, \ref{eq:magnetic-bc}) in a matrix form,
\begin{align}
    \begin{pmatrix}
    1 & -1\\
    \sqrt{\epsilon_+} & \sqrt{\epsilon_-}
    \end{pmatrix}
    \begin{pmatrix}
        E_x^+(\omega)
        \\
        E_x^-(\omega)
    \end{pmatrix}
    =
    \begin{pmatrix}
        0
        \\
        Z_0 J_x(\omega)
    \end{pmatrix}.
\end{align}
Inverting, we can find
\begin{align}
E_x^\pm(\omega)
&= \frac{Z_0 J_x(\omega)}{\sqrt{\epsilon_+}+\sqrt{\epsilon_-}},
\quad
H_y^\pm(\omega)
= \frac{\pm\sqrt{\epsilon_\pm}J_x(\omega)}{\sqrt{\epsilon_+}+\sqrt{\epsilon_-}},
\end{align}
The radiation intensities in the upper and lower media are calculated from the Poynting vector as following:
\begin{align}
    \frac{\mathrm{d}w_\pm}{\mathrm{d}\omega} 
    &= 
    \frac{\pm1}{2} \operatorname{Re} 
    \left[\{E_x^\pm(\omega)\}^* H_y^\pm(\omega)\right],
    \\
    &=
    \frac{1}{2} \frac{\sqrt{\epsilon_\pm}}{\sqrt{\epsilon_+}+\sqrt{\epsilon_-}}
    Z_0 |J_x(\omega)|^2.
\end{align}
This is the radiation intensity between an frequency interval $[\omega,\omega+\mathrm{d}\omega]$.
By integrating this quantity, we can obtain averaged radiation intensity.
Remind that the integration over the frequency is equivalent to taking the average in the time domain (Eq.~\eqref{radiation_intensity} in the main text).

\bibliography{ref}

\end{document}